# A High Accuracy and High Sensitivity System Architecture for Electrical Impedance Tomography System

Hui Li, Boxiao Liu, Yongfu Li, Guoxing Wang and Yong Lian
Department of Micro-Nano Electronics, Shanghai Jiao Tong University
Shanghai, China
E-mail: elelihui@sjtu.edu.cn

*Abstract*— A high accuracy and high sensitivity system architecture is proposed for the read-out circuit of electrical impedance tomography system-on-chip. The switched ratio-metric technique is applied in the proposed architecture. The proposed system architecture minimizes the device noise by processing signals from both read-out electrodes and the stimulus. The quantized signals are post-processed in the digital processing unit for proper signal demodulation and impedance ratio calculation. Our proposed architecture improves the sensitivity of the read-out circuit, cancels out the gain fluctuations in the system, and overcomes the effects of motion artifacts on measurements.

## I. INTRODUCTION

Acute lung injury is a very common complication in intensive care unit [1]. It has been demonstrated that mechanical ventilation with ventilator settings which do not suit the individual requirements of the diseased lung can injure the cellular structures of the lung tissue. Lung protective ventilation may prevent mechanical pulmonary injury from finding optimal PEEP (Positive End-Expiratory Pressure) and tidal volume settings for the individual patient. However, such task remains a constant challenge in clinical practice [2]. Although Computed Tomography (CT) provides detailed regional information about the lung, it is not suitable for continuous regional lung monitoring at the bedside and thus cannot be used to guide the routine adjustment of ventilator settings.

Compared with CT, Electrical Impedance Tomography (EIT) can achieve non-invasive, radiation-free monitoring of regional lung function. It is a medical imaging technique based on the electrical properties (resistivity, permittivity) of tissues and organs. The changes in tissue electrical properties are used to reflect the volume changes of organs. By measuring bio-impedance around the torso, EIT can provide dynamic images of lung, which may effectively prevent mechanical pulmonary injury. As EIT depending on the very small variations of impedance due to the change of air volume in the lung, the accuracy and sensitivity of read-out circuit in the EIT system are the critical parameters in enhancing the performance.

Fig. 1 shows the lung EIT system with 16 electrodes. An alternating current (AC) is injected into the region between two adjacent electrodes, and voltage signals are measured on all other electrodes in turn, i.e. 13 channels for voltage recording across the chest. Sequential pairs are then successively used for injecting current until all possible combinations have been measured. Finally, a complete data set of 208 (13 × 16) combinations is collected. The measurements could be utilized to recover one frame tomographic image of lung.

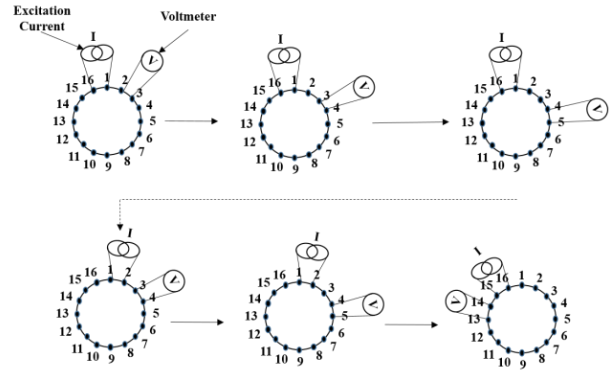

Figure 1. Electrode organization of 16-electrode EIT system.

Several EIT systems were developed for ventilation monitoring [3-4]. However, these traditional EIT systems were constructed with commercial off-the-shell (COTS) discrete components introducing noise through the wires during the signal transmission from electrodes to processing system. With the maturity of integrated circuit technology, system-on-chip (SoC) has contributed to portable EIT systems [5-6] which are smaller in size, low cost, low power and feasible for home use. The present EIT systems have much room for improvement for the sensing accuracy. Present systems adopted the I/Q demodulation technique in the analogue part to reduce the power consumption of analog-to-digital converter (ADC). The phase-sensitive signals pass through a low-pass filter (LPF) to derive the direct current (DC) value. However, it is not practical to implement an on-chip high-order analog low-pass filter to achieve a high degree of measurement accuracy [5]. The present systems also suffer from temperature variation and gain fluctuations. In addition, it is difficult to realize a high sensitivity system with a limited resolution ADC. Though the resolution of ADC is limited, M. Kim, *et al*, propose to average the measurements in different cycles to improve the sensitivity [6].

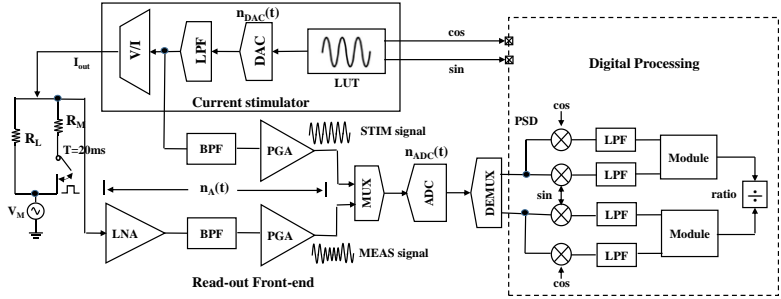
Figure 2. The block diagram of the proposed digital lock-in amplifier.

In this paper, a digital lock-in amplifier is proposed to increase the accuracy and sensitivity of the EIT system. Unlike the present systems, the proposed architecture quantizes the stimulus signals directly, avoiding the high-order analog low-pass filter. The digital lock-in amplifier combines with the switched ratio-metric technique to achieve a more accurate voltage measurement. The sensitivity can be improved by increasing the sampling rate of ADC rather than the resolution.

The rest of the paper is organized as follows. In Section II, the proposed system architecture and functional blocks are introduced. Section III shows the simulation results, and finally Section IV concludes the paper.

## II. PROPOSED ARCHITECTURE

The block diagram of proposed system architecture based on digital lock-in amplifier is shown in Fig. 2. It consists of three main functional blocks, namely current stimulator, read-out front-end and digital processing unit. The current stimulator generates the stimulus current as well as the reference signals for digital processing unit. The read-out front-end plays a key role in acquiring the signals, which consists of low noise amplifier (LNA), band-pass filter (BPF), programmable gain amplifier (PGA), analog multiplexer (MUX), ADC and digital de-multiplexer (DEMUX). The filters serve as an anti-aliasing filter and suppress the motion artifacts and intrinsic noise fluctuations. The PGA adjusts the signals' amplitude from various electrode pairs and provides the DC operating point for the signals. A time-interleaved ADC is proposed to sample the input signals alternatively. The output data of ADC is transferred to digital processing unit which performs phase-sensitive detection, signal-filtering and ratio calculation.

### A. Human Model

There are only subtle amplitude changes in tissue electrical properties due to ventilation of lung. The voltage changes at the surface of a human body is less than 100 μV. It can be inferred that the resistance difference of tissue is at mΩ level between inhalation and exhalation. Therefore, we use two parallel resistors ($R_L$ and $R_M$) to model the resistance difference during inhalation and exhalation, as shown in Fig. 2 (the left part). $R_M$ is several thousand times larger than $R_L$. The switching operation helps to simulate the respiratory process. Additional $V_M$ source can be regarded as the source of motion artifacts.

### B. Current Stimulator

The excitation signal generated by current stimulator is based on a sinusoidal signal stored in the look-up table (LUT). The sinusoidal signal acts as a reference for the digital processing unit to calibrate the errors in the system. For reproducing of digitally coded sinusoidal signal, the digital-to-analog converter (DAC) needs to have the same resolution as the ADC, and its data conversion rate is half that of the ADC. Existing ADC [7] and DAC [8] can be used for the current stimulator. A $2^{nd}$ order LPF is included to reduce the harmonic distortion of DAC's output signal.

### C. Read-out Front-end

To maximize the dynamic range of the ADC, it is preferable to amplify the changes of the input signal rather than the overall input signal. As the input signal contains the changes of impedance, the injected reference signal, and the DC offset from electrode, it is important to set the overall front-end gain to a proper value to prevent the output from saturation. Considering the magnitude of injected signal and possible motion artifacts, we set the gain of LNA to 20 dB and PGA to 0-20 dB. The center frequency of BPF is adjustable in accordance with the specific frequency of stimulus signals. As a result, the effect of motion artifacts on measurements will have less impact on the impedance.

The noise poses a fundamental limit to the accuracy of EIT system. Although the interference of additive noise can be suppressed by phase-sensitive detection, the effects of the amplitude noise cannot be reduced effectively. The amplitude noise is multiplicative noise which can modulate the amplitude of the signal randomly. We refer these noises as the gain fluctuation in the system, which exist in both reference generation stage and signal acquisition chain. For the generation stage, the unavoidable 1/f noise of the reference voltage used by DAC can cause the output voltage changing which leads to a fluctuated excitation signal directly. In the acquisition chain, similar effects can be found in ADC. The output of ADC is modulated randomly by comparing the input signal with the fluctuated reference which affects the measurement result directly. In the analog stages, the resistors setting the gain of amplifiers are the noise sources.

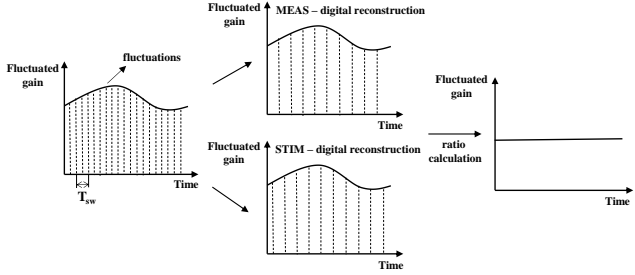
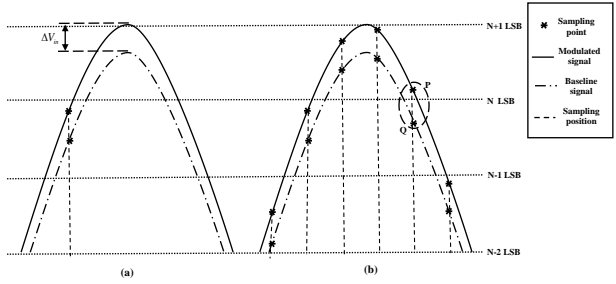

Figure 3. The process of cancelling out fluctuation gain.

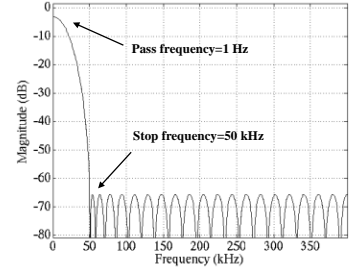

Figure 5. The frequency response of FIR.

Figure 4. Illustration of sensitivity enhancement by increasing sampling rate.

In the EIT system, the subtle voltage changes due to ventilation of lung are also modulated on the stimulus signal. It is difficult to separate the impedance information from the amplitude noise by filtering because the measured result is mixed with impedance signal and gain fluctuations. The system obtains poor accuracy measurements without cancelling out the gain fluctuations, not to mention the sensitivity.

The effects of these gain fluctuations in proposed architecture are illustrated in Fig. 2, where the $n_{DAC}(t)$, $n_{ADC}(t)$ (modulated function) and $n_A(t)$ (gain function) represent the noise from DAC, ADC, and analog stages. Supposing the $G_1$ is the total gain of BPF and PGA for STIM signal, the $G_2$ is the total gain of LNA, BPF and PGA for MEAS signal, then the amplitudes of the two signals can be described as (1) and (2).

$$A_{STIM} = A \times n_A(t) \times (1+n_{DAC}(t)) \times (1+n_{ADC}(t)) \times G_1 \quad (1)$$

$$A_{MEAS} = A \times |F_{MEAS}| \times n_A(t) \times (1+n_{DAC}(t)) \\ \times (1+n_{ADC}(t+T_{sw})) \times G_2 \quad (2)$$

where $A$ is the amplitude of the stimulus voltage, $|F_{MEAS}|$ is the magnitude of load transfer function which is the one we are measuring.

To obtain $|F_{MEAS}|$ from (1) and (2), ratio-metric technique is adopted to minimize the gain fluctuations [9], which is illustrated in Fig. 3. In the proposed read-out front-end, a single ADC is used to sample the STIM signal and MEAS signal alternatively. The MUX switching frequency is chosen fast enough to assume the same fluctuated gain for the two signals' samples in a period. The fluctuated gain can affect both STIM and MEAS signals in the same way when $T_{sw} \ll T_{fluctuations}$ is guaranteed. Here, $T_{fluctuations}$ refers to the period of the gain fluctuations. After performing the ratio calculation between the amplitudes of MEAS signal and STIM signal in (3), the fluctuated gain can be kept as a nearly constant value. As a result, according to (3), the magnitude of load transfer function $|F_{MEAS}|$ can be extracted precisely.

$$\frac{A_{MEAS}}{A_{STIM}} = |F_{MEAS}| \times \frac{1+n_{ADC}(t+T_{sw})}{1+n_{ADC}(t)} \times G_D \quad (3)$$

where $t$ follows $T_{fluctuations}$, $G_D$ is the gain difference between the two paths and $T_{sw}$ is the MUX switching period.

In terms of the sensitivity, increasing ADC resolution definitely helps. However, a high resolution ADC is difficult to design, and most of them consume more power and silicon area. An alternative way of boosting the sensitivity without the use of high resolution ADC is to increase the sampling frequency. Let's illustrate this by using Fig. 4. Assume that switching on and off $R_M$ caused a change in the input signal by $\Delta V_{in}$, which is smaller than 1 Least Significant Bit (LSB) of the ADC. Note that switching on and off $R_M$ can also be viewed as performing the modulation on a baseline signal. As Fig. 4(a) illustrates, at the Nyquest sampling frequency, the sampling points for both modulated and baseline signals fall within the same LSB interval of the ADC. In such a case, it is not possible to differentiate between modulated signal and baseline signal by the ADC, i.e. the ADC is not able to detect the change in the input. If we increase the sampling frequency of ADC, more sampling points are presented in a period, as is shown in Fig. 4(b). It can be seen that the sampling points (Q and P) of baseline signal and modulated signal are separated in two different LSB intervals. As a result, the modulated information can be captured by the ADC. In other words, by increasing the sampling frequency, it is likely to capture the smaller changes in the signal. Therefore, the sensitivity can be improved by increasing sampling frequency rather than the resolution of ADC.

### D. Digital Processing

The digital processing unit uses a digital-based phase-sensitive demodulation system and several 24[th]-order FIR LPFs to distinguish and reconstruct the digitized samples into STIM and MEAS signals. Fig. 5 shows the frequency response of the FIR filter when the frequency of stimulus current is 50 kHz. The pass frequency is set as 1 Hz for actual respiratory process. The gain of the filter outside pass band is lower than -60 dB which is low enough to extract the DC signal from high frequency components. The rest processing unit simply calculates the ratio of amplitudes of the MEAS and STIM signals to remove the gain fluctuations.

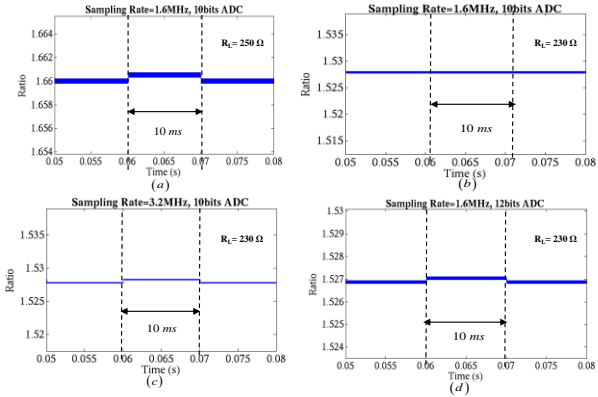

Figure 6. The detected results of "small change": (a) The detected small change when $R_M$ is switched on and off; (b) $R_L$ is changed to 230 Ω compared with (a); (c) The sampling rate is increased to 3.2 MHz compared with (b); (d) The resolution is increased to 12-bit compared with (b).

## III. SIMULATION RESULTS

The system is validated through Cadence® Virtuoso® and Matlab® co-simulation model. The analog circuits are implemented in the Cadence Virtuoso AMS environment while the digital signals are post-processed in the MATLAB environment.

The supply voltage for the analogue part is 2 V. The amplitude of STIM signal is 500 mV. As for the load resistance, $R_L$ is 250 Ω and $R_M$ is 1 MΩ. This results a change of resistance by 62.5 mΩ when $R_M$ is connected in parallel with $R_L$ through a periodic switch, as shown in Fig. 2. The switching period is set to 20 ms for this simulation. The amplitude of $V_M$ source imitating motion artifacts is 5 mV. For lung tissue conductivity difference, the most obvious deviation appears in the region of tens of kHz [10]. Therefore, the frequency of the stimulus current is set to 50 kHz and amplitude is set to 300 μA which contributes to a 75 mV baseline voltage for MEAS signal. By switching on $R_M$, a voltage change of 18.75 μV is introduced at the input of read-out front-end periodically. The small voltage change is amplified up to 187.5 μV which is still less than 1 LSB (1.95 mV) of a 10-bit ADC. By using the proposed system architecture and set the sampling rate of ADC to 1.6 MHz, we can capture the small change at the input as shown in Fig. 6(a). This verifies that the proposed system architecture achieves much higher accuracy than that of the ADC resolution.

To verify the sensitivity improvement with the increase of sampling frequency and ADC resolution, the resistance of $R_L$ is changed to 230 Ω. By switching on $R_M$, 52.9 mΩ change of resistance is introduced. After adjusting $R_L$, the "small change" caused by switching on $R_M$ is no longer detectable, as illustrated in Fig. 6(b). Reducing $R_L$ makes the "small change" even smaller so that higher sensitivity is needed. To enhance the sensitivity, we increase the sampling rate to 3.2 MHz. From Fig. 6(c), it is obvious that the "small change" becomes visible. By increasing the ADC resolution from 10-bit to 12-bit while keeping the sampling rate as 1.6 MHz, we can detect the "small change" too, as shown in Fig. 6(d). The simulation results demonstrate that the sensitivity can be improved by increasing either ADC resolution or sampling rate.

## IV. CONCLUSION

This paper presents a high accuracy and high sensitivity system architecture with switched ratio-metric signal-processing technique for electrical impedance tomography system. From our experimental result, a minimum sensitivity of 62.5 mΩ can be measured with proposed system when the sampling frequency is set to 1.6MHz. The sensitivity can be further improved by increasing the sampling rate and resolution of the ADC. To minimize the variability in the analog circuit for advanced CMOS technology, our proposed system is able to relax the analog requirement and able to take advantage in the digital circuit to implement the phase-sensitive detection.

ACKNOWLEDGMENT

This work has been supported by the National Key Research and Development Program of China, project number '2016YFE0116900'.